\documentclass[11pt]{4ssmmp}

\usepackage{latexsym}

\newcommand{\beq}{\begin{equation}}
\newcommand{\eeq}{\end{equation}}
\newcommand\be{\begin{equation} }
\newcommand\bea{\begin{eqnarray}}
\newcommand\ee{\end{equation}}
\newcommand\eea{\end{eqnarray}}

\def\Tr{{\rm Tr}\,}

\def\m{\mu}          \def\n{\nu}
       
       \def\r{\rho}
\def\s{\sigma}

         \def\G{\Gamma}

\def\ci{{\cal I}}      
      \def\cl{{\cal L}}

      \def\car{{\cal R}}

\usepackage{epsfig}

\usepackage{fancyhdr}    
\usepackage{graphicx}
\usepackage{rotating}

\begin{document}
 \baselineskip=11pt

\title{Signal for space-time noncommutativity: the $Z \to \gamma\gamma$ decay in
the renormalizable gauge sector of the $\theta$-expanded NCSM\hspace{.25mm}
\thanks{\,Based on presentation given at the IV Summer School in Modern 
Mathematical Physics, Belgrad, Serbia, September 3-14, 2006
and LHC Days in Split, Croatia, October 2-7, 2006.
Work supported by the Croatian Ministry of Science, Education and
Sport project 098-0982930-2900.
}}
\author{\bf{Josip Trampeti\' c}\hspace{.25mm}\thanks{\,e-mail address: josipt@rex.irb.hr}
\\ \normalsize{Rudjer Bo\v skovi\' c Institute, Zagreb, Croatia} \vspace{2mm} \\ 
}

\date{}

\maketitle

\begin{abstract}

We propose the $Z\rightarrow \gamma\gamma$
decay, a process strictly
forbidden in the standard model, as a signal 
suitable for the search of  noncommutativity
of coordinates at very short distances. 
We compute the $Z\rightarrow \gamma\gamma$
partial widthin the framework of the recently proposed 
renormalizable gauge sector of the noncommutative standard model.
The one-loop renormalizability is obtained
for the model containing the usual six representations
of  matter fields of the first generation.
Even more, the noncommutative part is finite or free of divergences, 
showing that perhaps new interaction symmetry exists
in the noncommutative gauge sector of the model. Discovery of such symmetry 
would be of tremendous importance in further search
for the violation of the Lorentz invariance at very high energies.
Experimental possibilities of $Z\rightarrow \gamma\gamma$ decay are analyzed and a
firm bound to the scale of the 
noncommutativity parameter is set around 1 TeV.
\end{abstract}

\clearpage 


Gauge theories can  be extended to a noncommutative (NC) setting
in  different ways. In our model, the classical action is obtained
via a two-step procedure. First, the noncommutative Yang-Mills 
(NCYM) is equipped with
a star product carrying information about the underlying
noncommutative manifold, and, second, the $\star$-product and
noncommutative fields are expanded in the noncommutative parameter
$\theta$ using the Seiberg-Witten (SW) map \cite{Seiberg:1999vs}. 
In this approach,
noncommutativity is treated perturbatively.  The major advantage
is that models with any gauge group and any particle content can
be constructed
\cite{Wess,Calmet:2001na,Blazenka,Aschieri:2002mc,Goran,Buric:2006nr}, 
so we can
construct the standard model (SM). 
Commutative  gauge symmetry is the underlying
symmetry of the theory and is present in each order of the
$\theta$-expansion. Noncommutative (NC) symmetry, on the other hand,
exists only in the full theory, i.e. after summation.

There are a number of versions of the noncommutative standard model
(NCSM) in the $\theta$-expanded approach,
\cite{Calmet:2001na,Blazenka,Aschieri:2002mc,Goran}. 
The action is gauge invariant;
furthermore, it has been proved that the action is anomaly free
whenever its commutative counterpart is also anomaly free
\cite{Brandt:2003fx}.
The argument of renormalizability was previously included in the
construction of field theories on noncommutative Minkowski space
producing not only the one-loop renormalizable model 
\cite{Buric:2005xe}, but the
model containing one-loop quantum corrections free of divergences
\cite{Buric:2006wm}, contrary to previous results
\cite{Wulkenhaar:2001sq,Maja}.

In \cite{Buric:2006wm} we analyzed the gauge theory based on
the $\rm  U(1)_Y \times SU(2)_L \times SU(3)_C$ group:
we succeeded in constructing a model which had the renormalizable
gauge sector to $\theta$-linear order. 
The condition of the gauge sector renormalizability determines the 
additional
$\theta$-linear interactions between gauge bosons.  

Experimental evidence for noncommutativity coming
from the gauge sector should be searched for
in the process of the $Z\rightarrow \gamma\gamma$ decay, kinematically
allowed for on-shell particles \cite{Buric:2006wm,Buric:2006nr}. As it is
forbidden in the SM by angular momentum conservation and Bose statistics
(Landau-Pomeranchuk-Yang Theorem), it
would serve as a clear signal for the existence of
space-time noncommutativity.
Signatures of noncommutativity were
discussed previously within particle physics in
\cite{Buric:2006nr,Josip,Ohl:2004tn}. 

The noncommutative space which we consider is the flat Minkowski space,
generated by four hermitian coordinates $\widehat x^\mu$
which satisfy the commutation rule
\begin{equation}
[\widehat x^\mu,\widehat x^\nu ] =i\theta^{\mu\nu}={\rm const}.
\label{Mink}
\end{equation}
The algebra of the functions $\widehat\phi (\widehat x)$,
$\widehat \chi (\widehat x)$
on this space can be represented by the algebra of the functions
$\widehat \phi (x)$, $\widehat \chi (x)$ on the commutative
$ {\bf R}^4$ with
the Moyal-Weyl multiplication:
\begin{equation}
\label{moyal} \widehat\phi (x)\star \widehat\chi (x) =
      e^{\frac{i}{2}\,\theta^{\mu\nu}\frac{\partial}{\partial x^\mu}
      \frac{\partial}{ \partial
      y^\nu}}\widehat \phi (x)\widehat \chi (y)|_{y\to x}\ .
\end{equation}
It is possible to represent the action of an arbitrary Lie group $G$
(with the generators denoted by $T^a$) on noncommutative space.
In analogy to the ordinary  case,
one introduces the gauge parameter $\widehat \Lambda ( x)$
and the vector potential $\widehat V_\mu( x)$.  The main difference
is that the noncommutative
$\widehat \Lambda$ and $\widehat V_\mu $  cannot take values
in the Lie algebra $\cal G$ of the group $G$:  they are
enveloping algebra-valued.
The noncommutative gauge field strength  $ \widehat F_{\mu\nu}$ is 
\begin{equation}
\widehat F_{\mu\nu} = \partial_\mu\widehat V_\nu -
\partial_\nu \widehat V_\mu - i(\widehat V_\mu\star\widehat V_\nu -
\widehat V_\nu\star\widehat V_\mu ).                         
\label{f}
\end{equation}
There is, however, a relation between the noncommutative
gauge symmetry and the commutative one:
it is given by the Seiberg-Witten (SW) mapping \cite{Seiberg:1999vs}.
Namely, the matter fields
 $\widehat \phi$, the gauge fields  $\widehat V_\mu$, $\widehat F_{\mu\nu}$
 and the gauge parameter
$\widehat\Lambda$ can be expanded in the noncommutative  $\theta^{\mu\nu}$
and in the commutative  $V_{\mu}$ and $F_{\mu\nu}$.
 This expansion  coincides with the expansion in
the generators of the enveloping algebra of $\cal G$,
$\{ T^a$, $:T^aT^b:$, $:T^aT^bT^c: \}$;
 here \ $:\ :$ \ denotes the symmetrized product.
The SW map is obtained as a solution to the gauge-closing condition
of infinitesimal (noncommutative) transformations.
The expansions of the NC vector potential and of the field strength,
up to first order in $\theta$, read
\begin{eqnarray}
&&
\widehat V_\rho(x) =V_\rho(x) -\frac 14 \,\theta ^{\mu\nu}\left\{ V_\mu(x),
\partial_\nu V_\rho(x) +F_{\nu \rho}(x)\right\}
+\dots  ,
\label{expansion}\\
&&
\widehat F_{\rho\sigma} =   F_{\rho\sigma} +\frac{1}{4}\theta^{\mu\nu}
\Big(2\{F_{\mu\rho},F_{\nu\sigma}\} -
\{ V _\mu ,(\partial_\nu +D_\nu )F_{\rho\sigma} \}\Big)
+\dots ,
\label{FF}
\end{eqnarray}
where $D_{\nu}= \partial_{\nu}~-i[V_{\nu},~~~]$ 
is the commutative covariant derivative.

The solution for the SW map given above is  not unique 
and along with (\ref{FF}) all expressions ${\widehat V}^\prime_{\mu}$,
${\widehat F}^\prime_{\mu\nu}$ of the form
\begin{equation}
{\widehat V}^\prime_{\mu} = {\widehat V}_{\mu} +  X_\mu, \quad
{\widehat F}^\prime_{\mu\nu} = {\widehat F}_{\mu\nu} + D_\mu X_\nu - D_\nu X_\mu
\label{nonuniq}
\end{equation}
are solutions to the closing condition to linear order, if
 $X_\mu $ is a gauge covariant expression linear in $\theta$, otherwise
arbitrary. One can think of this transformation as of a redefinition of
the fields $V_\mu$ and $F_{\mu\nu}$.

Taking the action of the  noncommutative gauge theory, analogous to
that of the ordinary Yang-Mills theory with
the commutative field strengths replaced by the noncommutative ones,
\begin{equation}
S=-\frac{1}{2}\Tr \int d^4x\,\widehat F_{\mu\nu}\star\widehat F^{\mu\nu} ,
\label{action}
\end{equation}
and expanding the fields as in (\ref{expansion}-\ref{FF})
and the $\star$-product in $\theta$, we obtain the expression
\begin{equation}
S =-\frac{1}{2}\Tr\int d^4x\,F_{\mu\nu}F^{\mu\nu}+\theta^{\mu\nu}\,\Tr\int d^4x\,
\Big(\frac 14 F_{\mu\nu}F_{\rho\sigma}-
F_{\mu\rho}F_{\nu\sigma} \Big)F^{\rho\sigma},
\label{act}
\end{equation}
which is  the starting point for the analysis of $\theta$-expanded
noncommutative gauge models. 
Due to the renormalizability condition, 
we add term, including NC freedom parameter 
$\frac{1}{4}(a-1)$, to the original Lagrangian,
producing the following general form of
the noncommutative gauge field action:
\begin{equation}
S =-\frac 12\Tr\int d^4x\,F_{\mu\nu}F^{\mu\nu}+\theta^{\mu\nu}\,\Tr\int d^4x\,
\big(\frac a4 \, F_{\mu\nu}F_{\rho\sigma}-
F_{\mu\rho}F_{\nu\sigma} \big)F^{\rho\sigma} .
\label{Act}
\end{equation}

The most general form of the NC action, invariant under the NC gauge transformation, is
given in \cite{Calmet:2001na,Aschieri:2002mc,Goran,Blazenka},
\begin{eqnarray}
S_{\rm gauge} = -\frac{1}{2}  \int
d^4x \, \sum_{\cal R} {C_{\cal R}} {\Tr}\Big(
{\cal R}(\widehat F_{\mu \nu}) \star {\cal R}(\widehat F^{\mu
\nu})\Big)\,.
\label{action1}
\end{eqnarray}
The sum in (\ref{action1}) is, in principle, taken  over all irreducible representations
${\cal R}$  of $\rm G_{SM}$  with arbitrary weights $C_{\cal R}$.
Obviously, gauge models are representation dependent  in the NC case:
the choice of  representations
has a strong influence on the theory, on both the form of interactions and
the renormalizability properties.

Expanding the NC gauge action (\ref{action1}) 
to first order in the noncommutativity parameter $\theta$, we obtain
\begin{eqnarray}
&&S_{\rm gauge} =-\frac{1}{2}\sum_{\car}C_\car\Tr\int d^4x\,\car(F_{\mu\nu})\car(F^{\mu\nu})
\label{act1} \\
&&+\theta^{\mu\nu}\,\sum_{\car}C_\car\Tr\int d^4x\,
\Big(\frac a4 \,\car(F_{\mu\nu})\car(F_{\rho\sigma})
-\car(F_{\mu\rho})\car(F_{\nu\sigma}) \Big)\car(F^{\rho\sigma}).
\nonumber
\end{eqnarray}
The arbitrariness in the gauge action, 
introduced through the coefficient $a$, reflects
in part also the nonuniqueness of the SW map. 
As we have already mentioned, 
renormalizability points out the value $a=3$ as physical; however,  
we keep the value of $a$ arbitrary in calculations 
and use $a=3$ at the end.

Note that by generalizing the expression (\ref{FF}) to
equivalent form
\begin{eqnarray}
{\widehat F}_{\mu\nu}(a)&=&F_{\mu\nu} 
+ \frac{1}{4}\theta^{\rho\tau}
\Big(2\{F_{\mu\rho},F_{\nu\tau}\}-a\{V_{\rho},(\partial_{\tau}
+D_{\tau})F_{\mu\nu}\}\Big)\,,
\label{fieldsa}            
\end{eqnarray}
one could also obtain 
the actions (\ref{Act},\ref{act1}) directly from 
(\ref{action},\ref{action1}).\footnote{This is in part due to the properties of 
the integral over the two-function $\star$-product, i.e. the Stokes theorem.} 
The important question, if the freedom parameter $a$ is eventually comming from 
different class of SW maps and/or some other new interaction symmetry
extends the purpose of this presentation and, consequentlly,
shall be discussed elsewhere.

The noncommutative correction, that is the $\theta$-linear part of the
Lagrangian, reads\bea
\cl^\theta &=& \sum \cl^\theta _i =g^{\prime 3}\kappa_1\theta^{\m\n}
\left( \frac a4 f_{\m\n}f_{\r\s}f^{\r\s}-f_{\m\r}f_{\n\s}f^{\r\s}\right)
\nonumber\\
&+&g^3 \kappa_4^{ijk}\theta^{\m\n}
 \left(\frac a4 B_{\m\n}^i B_{\r\s }^j B^{\r\s k} - B_{\m\r}^i B_{\n\s }^j
B^{\r\s k} \right) \nonumber\\
&+& g^3_S\kappa_5^{abc}\theta^{\m\n}
 \left( \frac{ a}{4}G_{\m\n}^a G_{\r\s }^b G^{\r\s c}- G_{\m\r}^a G_{\n\s }^b
G^{\r\s c}\right) \nonumber \\
&+& g^\prime g^2  \kappa_2 \theta^{\m\n}
\left(\frac a4 f_{\m\n}B_{\r\s }^i B^{\r\s i} -f_{\m\r}B_{\n\s }^i B^{\r\s i} +
c.p.\right) \nonumber \\
&+& g^\prime g^2_S  \kappa_3 \theta^{\m\n}
\left(\frac a4 f_{\m\n}G_{\r\s }^a G^{\r\s a} -f_{\m\r}G_{\n\s }^a G^{\r\s a}
+c.p. \right),\label{lterm}
\eea
where the $c.p.$ in (\ref{lterm}) denotes the addition of the terms
obtained by a cyclic permutation of
fields without changing the positions of indices.
Here, $f_{\mu\nu}$, $B^i_{\mu\nu}$, and $G^b_{\mu\nu}$
are the physical field strengths which correspond
to  $\rm U(1)_Y$, $\rm SU(2)_L$, and $\rm SU(3)_C$, respectively.
The couplings $\kappa_i,\;(i=1,...,5)$, 
as functions of the weights $C_\car$, that is of the $C_i(=1/g_i^2),\; i=1,...,6$,
are parameters of the model.
The couplings in (\ref{lterm}) are defined as follows:
\bea
\kappa_1 &=&\sum_{\car}C_{\car}
d(\car_2)d(\car_3)\car_1(Y)^3,
\label{k1} \\
\kappa_2 \delta^{ij} &=& \sum_{\car}C_{\car}
d(\car_3)\car_1 (Y)\Tr (\car_2(T^i_L)\car_2(T^j_L)),
\label{k2} \\
\kappa_3 \delta^{ab}&=& \sum_{\car}C_{\car}
d(\car_2)\car_1 (Y)\Tr (\car_3(T^a_S)\car_3(T^b_S)),
\label{k3} \\
\kappa_4 ^{ijk}&=& \frac12\sum_{\car}C_{\car}
d(\car_3)\Tr (\{\car_2(T^i_L),\car_2(T^j_L)\}\car_2(T^k_L)),
\label{k4} \\
\kappa_5^{abc} &=&\frac12 \sum_{\car}C_{\car}
d(\car_2)\Tr (\{\car_3(T^a_S),\car_3(T^b_S)\}\car_3(T^c_S)) .
\label{k5}
\eea
The $\kappa_1, \dots  ,\kappa_5$ depend on the representations of matter fields
through the dependence on the coefficients $C_\car$.
For the first generation of the standard model there are six such
representations, summarized in Table 1 of \cite{Blazenka}; they  produce six
independent constants $C_\car$ \footnote{We assume that $C_\car >0$;
therefore the six $C_\car$'s were denoted by
$\frac{1}{g_i^2}\,,i=1,...,6$, in \cite{Calmet:2001na,Goran}.}.
However, one can immediately verify that $\kappa_4^{ijk} = 0$.
This follows from the fact that the symmetric coefficients
$d^{ijk}$ of $\rm SU(2)$ vanish for all irreducible representations. In
addition, we take that $\kappa^{abc}_5 = 0$. The argument for this assumption is
related to the invariance of the color sector of the SM  under charge
conjugation. Although apparently in Table 1 from \cite{Blazenka} 
one has only the fundamental
representation {\bf 3} of $\rm SU(3)_C$, there are in fact both ${\bf 3}$ and
${\bf \bar 3}$ representations with the same weights, $C_{{\bf 3}} =C_{{\bf\bar
3}}$. In the Lagrangian this corresponds to writing each minimally-coupled
quark term as a half of the sum of the original  and the charge-conjugated terms.
Since the symmetric coefficients for the {\bf 3} and ${\bf \bar 3}$
representations satisfy $d^{abc}_{{\bf \bar 3}} = - d^{abc}_{{\bf 3}}$,
we obtain
\begin{equation}
 \kappa_5^{abc} = C_{{\bf 3}}d_{\bf 3}^{abc} + C_{{\bf \bar 3}}d_{\bf \bar
3}^{abc} =0 .\label{Kappa}
\end{equation}
We are left only with three nonvanishing couplings, $\kappa_1$,
$\kappa_2$, and $\kappa_3$, depending on six constants $C_1,\dots,C_6$:
\bea \kappa_1&=&-\,C_1-\frac14\,C_2+\frac89
\,C_3-\frac19 \,C_4+ \frac{1}{36} \,C_5+\frac14 \,C_6\,,
\nonumber \\
\kappa_2&=&-\,\frac14\,C_2+\frac14 \,C_5+\frac14 \,C_6\,;\;\;
\kappa_3= \,\frac13 \,C_3- \frac16\,C_4+ \frac16 \,C_5\,.
\label{kappa}
\eea
There are three relations among $C_i$'s:
\bea
\frac{1}{g^{\prime 2}}&=&\,2\,C_1+C_2+\frac83 \,C_3+\frac23 \,C_4+
\frac13 C_5+C_6\,,
\nonumber \\
\frac{1}{g^{2}}&=&\,C_2+3\,C_5+\,C_6\,;\;\;
\frac{1}{g^2_s}= \,C_3+ \,C_4+2 \,C_5\,,
\label{C}
\eea
in effect representing three consistency conditions imposed on
(\ref{act}) in a way to match the SM action at zeroth order in $\theta$.
See detailes in \cite{Goran}.
\begin{figure}
 \resizebox{.45\textwidth}{!}{\includegraphics{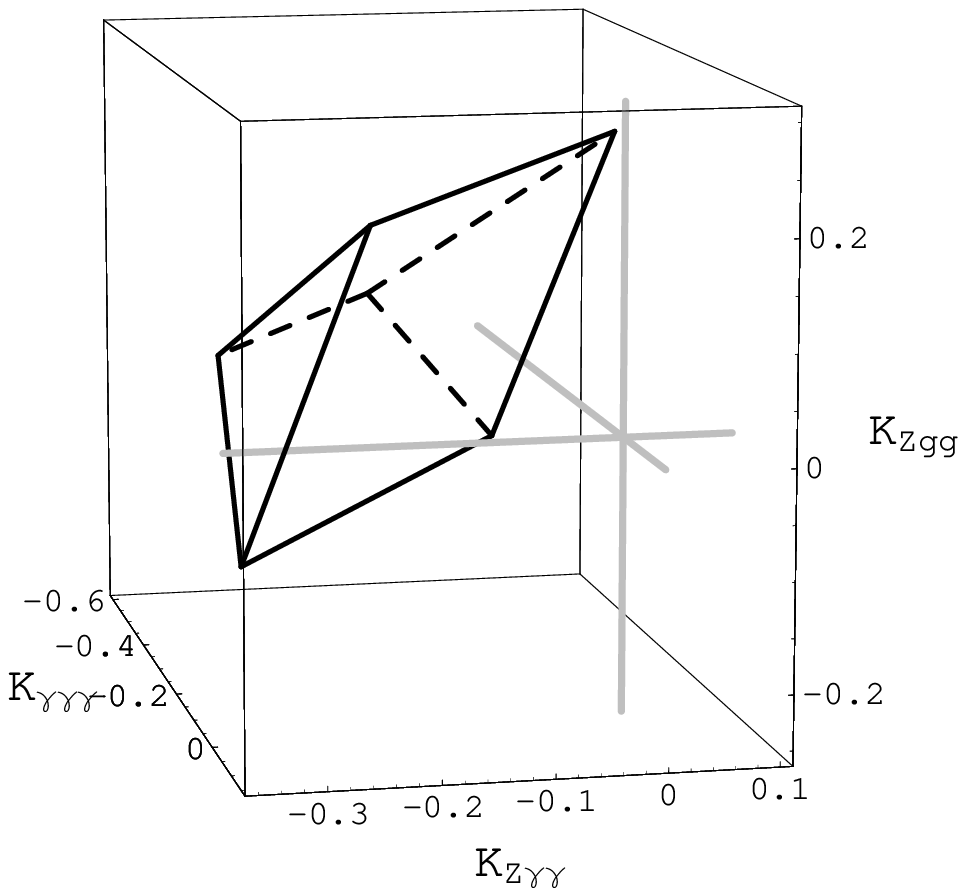}}
  \hspace{.2cm}
  \resizebox{.5\textwidth}{!}{\includegraphics{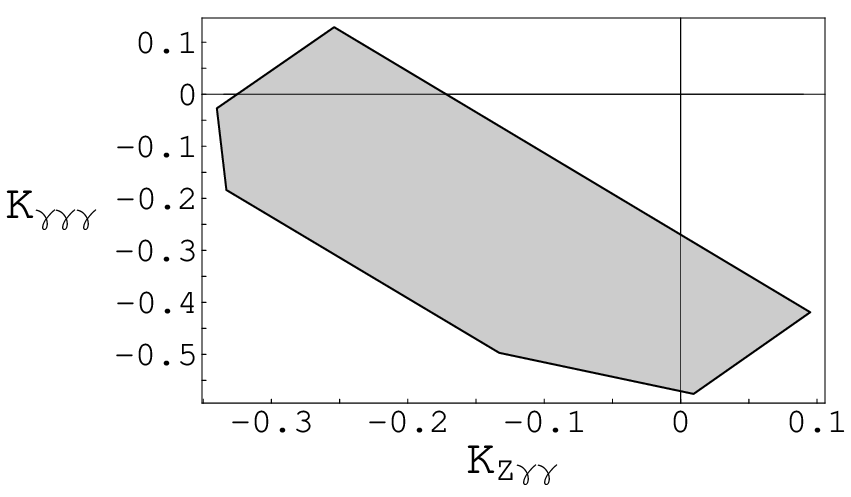}}
\caption{(a) The three-dimensional simplex that bounds possible values
 for the coupling constants ${\rm K}_{\gamma\gamma\gamma}$, 
 ${\rm K}_{Z\gamma\gamma}$ and ${\rm K}_{Zgg}$ at the  $M_Z$ scale. 
 The vertices of the simplex are:
 ($-0.184$,\,$-0.333$,\,$0.054$), ($-0.027$,\,$-0.340$,\,$-0.108$),
 ($0.129$,\,$-0.254$,\,$0.217$),
 ($-0.576$,\,$0.010$,\,$-0.108$), ($-0.497$,\,$-0.133$,\,$0.054$), and 
 ($-0.419$,\,$0.095$,\,$0.217$).
(b) The allowed region for ${\rm K}_{Z\gamma\gamma}$ 
 and ${\rm K}_{\gamma\gamma\gamma}$  at the $M_Z$ scale, projected from
 the simplex given in Fig (a). 
 The vertices of the polygon are:
 $(-0.333,\, -0.184)$, $(-0.340,\, -0.027)$, $(-0.254,\, 0.129)$, $(0.095,\, -0.419)$, 
 $(0.0095, \,-0.576)$, and $(-0.133,\, -0.497)$.
}
 \label{fig1}
\end{figure}

Fig.(\ref{fig1}) shows the three-dimensional simplex that bounds 
allowed values for the dimensionless coupling constants
${\rm K}_{\gamma\gamma\gamma}$, ${\rm K}_{Z\gamma\gamma}$ 
and ${\rm K}_{Zgg}$. For any choosen point within the simplex 
in Fig.(\ref{fig1}) the remaining coupling constants
${\rm K}_{Z Z \gamma}$, ${\rm K}_{Z Z Z}$, ${\rm K}_{WW \gamma}$, ${\rm K}_{WWZ}$ 
and ${\rm K}_{\gamma g g}$ are uniquely fixed by the NCSM \cite{Goran,Blazenka}.
This is true for any combination of three coupling constants.

Our total classical action reads
\bea 
S_{cl} &=& S_{SM} + \sum_{i=1}^3
S^{\theta}_i = g^{\prime
3}\kappa_1\theta^{\m\n} \int d^4 x\left( \frac a4
f_{\m\n}f_{\r\s}f^{\r\s}-f_{\m\r}f_{\n\s}f^{\r\s}\right)\nonumber\\
&+& g^\prime
g^2  \kappa_2 \theta^{\m\n}\int d^4 x\left(\frac a4 f_{\m\n}B_{\r\s }^i B^{\r\s
i} -f_{\m\r}B_{\n\s }^i B^{\r\s i} + c.p.\right) \nonumber \\[4pt]
&+& g^\prime g^2_S
\kappa_3 \theta^{\m\n}\int d^4 x\left(\frac a4 f_{\m\n}G_{\r\s }^a G^{\r\s a}
-f_{\m\r}G_{\n\s }^a G^{\r\s a} +c.p. \right).
\label{St}
\eea
\smallskip
The term $S^{\theta}_1$ in (\ref{St})  is one-loop renormalizable to
linear order in $\theta$ \cite{Buric:2005xe} since the one-loop correction
to the $S^{\theta}_1$ is of the second order in $\theta$.
We need to investigate only the
renormalizability of the remaining $S^{\theta}_2$ and $S^{\theta }_3$ parts of
the action (\ref{St}).

To realize the one-loop renormalization of the gauge part
action (\ref{St}), we apply, as before
\cite{Buric:2005xe,Buric:2006wm}, the background field method
\cite{'tHooft:1973us,PS}.  As we have already explained the
details of the method in \cite{Maja}, here we only discuss the
points needed for this computation. The main contribution to the
functional integral is given by the Gaussian integral. However,
technically, this is achieved by splitting the vector potential
into the classical-background  plus the quantum-fluctuation parts,
that is, $\phi_V\to \phi_V + {\bf\Phi}_V$, and by computing the
terms quadratic in the quantum fields. In this way we determine
the second functional derivative of the classical action, which is
possible since our interactions (\ref{St})
are of the polynomial type. The quantization is performed by the
functional integration over the quantum vector field ${\bf
\Phi}_V$ in the saddle-point approximation around the classical
(background) configuration $\phi _V$.

First, an advantage of the background field method is the guarantee
of covariance, because by doing the path integral the local symmetry
of the quantum field ${\bf \Phi}_V$ is fixed, while the gauge
symmetry of the background field ${\phi}_V$ is manifestly preserved.

Since we are dealing with gauge symmetry, our Lagrangian (\ref{St})
is singular owing to its invariance under the gauge group.
Therefore, a proper quantization of (\ref{St}) requires the presence
of the gauge fixing term $S_{\rm gf}[\phi]$, i.e. the
Feynman-Fadeev-Popov ghost appears in the effective action
\begin{equation}
\Gamma [\phi] = S_{\rm cl}[\phi] + S_{\rm gf}[\phi] + \Gamma^{(1)}
[\phi],\qquad S_{\rm gf}[\phi]=-\frac 12 \int
\mathrm{d}^4x(D_{\mu}{\bf \Phi}_V^{\mu})^2\;.
 \label{GS2}
\end{equation}
The one-loop effective part $\Gamma^{(1)} [\phi]$ is given by
\begin{equation}
\Gamma ^{(1)}[\phi] =\frac{\mathrm{i}}{2}\log\det
S^{(2)}[\phi]=\frac{\mathrm{i}}{2}\Tr\log S^{(2)}[\phi].
\label{gama1}
\end{equation}
In (\ref{gama1}), the $ S^{(2)}[\phi]$ is the $2^{\rm
nd}$-functional derivative of the classical action, 
with the following structure: 
\begin{equation}
S^{2}=\Box+N_1+N_2+T_2+T_3+T_4 \ .
\label{Box}
\end{equation}
Here $N_1,N_2$ are
commutative vertices, while $T_2,T_3,T_4$ are noncommutative ones.
The indices denote the number of classical fields.  The one-loop
effective action computed by using the background field method is
\bea 
\G^{(1)}_{\theta,2}&=& \frac{\mathrm{i}}{2} \Tr \log
\left(\ci + \Box ^{-1} (N_1+N_2+T_2+T_3+T_4)\right)
\label{trlog}\\
&& \hspace*{-2cm}=\frac{\mathrm{i}}{2}\sum_{n=1}^\infty \frac{(-1)^{n+1}}{n} \Tr
\left(\Box^{-1}N_1+\Box ^{-1}N_2+\Box
^{-1}T_2+\Box^{-1}T_3+\Box^{-1}T_4 \right)^n\,. 
\nonumber 
\eea
As  the conventions and the notation are the same as in
\cite{Buric:2006wm}, we only encounter and discuss the final results.

The divergent one-loop vertex correction to (\ref{St})
as a function of the SW freedom parameter $a$ is \cite{Buric:2006wm}
\begin{eqnarray}
\Gamma_{\rm div}&=& \frac {11}{3 (4\pi)^2\epsilon}\int d^4 x
\Big(B_{\mu\nu}^iB^{\mu\nu i} + \frac{3}{2} G_{\mu\nu}^a G^{\mu\nu a} \Big)
\label{div}\\
&+&
\frac{4}{3(4\pi)^2\epsilon}{g^\prime} g^2  \kappa_2 (3-a)
\theta^{\mu\nu}\int d^4 x
\big(\frac{1}{4} f_{\mu\nu}B_{\rho\sigma}^i
- f_{\mu\rho}B_{\nu\sigma }^i \big) B^{\rho\sigma i}
\nonumber \\
&+& \frac{6}{3(4\pi)^2\epsilon}{g^\prime} g^2_S \kappa_3 (3-a)
\theta^{\mu\nu}\int d^4 x
\big(\frac{1}{4} f_{\mu\nu}G_{\rho\sigma }^a
- f_{\mu\rho}G_{\nu\sigma }^a \big)G^{\rho\sigma a}\,.
\nonumber 
\end{eqnarray}
From (\ref{div}) it is clear that the expanded gauge action 
(\ref{St}) is renormalizable
only for the value $a=3$ and,
its noncommutative part is finite or free of divergencies, so
the noncommutativity parameter $\theta$ need not be renormalized.
The results for the bare fields and couplings, are given in
\cite{Buric:2006wm}.

Note that we have also analized the renormalizability properties of 
the pure NC SU(N) gauge sector,
for vector fields in the adjoint representation \cite{Latas:2007eu}. 
We have found that this model is also renormalizable for $a=3$. 
However, to obtain renormalizability, we had to pay a price by necessity for the 
renormalization of the noncommutative deformation parameter $h$. 
In this way the parameter
$h$ and/or the scale of noncommutativity $\Lambda_{\rm NC}$ 
become running quantities, dependent on energy \cite{Latas:2007eu}. 

In addition, it was shown that the one-loop contributions to 
the U(1) gauge-field part of the noncommutative gauge theories in the 
enveloping-algebra formalism are renormalizable at first order in $\theta$ even if 
the scalar matter, with and without spontaneous symmetry breaking, contributions are taken
into account \cite{Martin:2006gw}. There is reasonable hope that 
the same conclusion should hold for SU(N), but 
the computations are expected to be extremely involving. 
Nevertheless, the results \cite{Martin:2006gw} further strengthen the philosophy  
which is embraced in our latest papers \cite{Buric:2006wm,Latas:2007eu}.

From the action (\ref{St}) we extract the
triple-gauge boson terms which are not present in the commutative SM Lagrangian.
In terms of the physical fields $A,\ W^{\pm},\ Z$, and $ G$ they are
\begin{eqnarray}
{\cal L}^{\theta}_{\gamma\gamma\gamma}&=&\frac{e}{4}\sin2{\theta_W}\;{\rm K}_{\gamma\gamma\gamma}
{\theta^{\rho\tau}}
A^{\mu\nu}\left(aA_{\mu\nu}A_{\rho\tau}-4A_{\mu\rho}A_{\nu\tau}\right),
\nonumber\\
{\rm K}_{\gamma\gamma\gamma}&=&\frac{1}{2}\; gg'(\kappa_1 + 3 \kappa_2);
\label{L1}\\
& & \nonumber \\
{\cal L}^{\theta}_{Z\gamma\gamma}&=&\frac{e}{4} \sin2{\theta_W}\,{\rm K}_{Z\gamma \gamma}\,
{\theta^{\rho\tau}}
\nonumber\\
&  \times& \left[2Z^{\mu\nu}\left(2A_{\mu\rho}A_{\nu\tau}-aA_{\mu\nu}A_{\rho\tau}\right)\right.
+\left. 8 Z_{\mu\rho}A^{\mu\nu}A_{\nu\tau} -aZ_{\rho\tau}A_{\mu\nu}A^{\mu\nu}\right],
 \nonumber\\
{\rm K}_{Z\gamma\gamma}&=&\frac{1}{2}\; \left[{g'}^2\kappa_1+\left({g'}^2-2g^2\right)\kappa_2\right];
\label{L2}
\end{eqnarray}
where $A_{\mu\nu} \equiv \partial_{\mu}A_{\nu}-\partial_{\nu}A_{\mu}$, etc.
The structure of the other interactions such as $ZZ\gamma$, $WWZ$, $ZZZ$, $Zgg$, 
and $\gamma gg$ is given in \cite{Blazenka,Goran}.

Next we focus on the branching ratio of the
$Z\rightarrow \gamma\gamma$ decay in the renormalizable model.
Note that each term from the
$\theta$-expanded action (\ref{St}),  (\ref{L1}) and  (\ref{L2}) is manifestly invariant
under the ordinary gauge transformations.
The gauge-invariant amplitude
${\cal A}^{\theta}_{Z\rightarrow \gamma\gamma}$ 
for the $Z(k_1)\rightarrow\gamma(k_2)\,\gamma(k_3)$ decay
in the momentum space reads
\begin{eqnarray}
{\cal A}^{\theta}_{Z\to \gamma\gamma}=-2e \sin2{\theta_W}{\rm K}_{Z\gamma \gamma}
{\Theta^{\mu\nu\rho}_3}(a;k_1,-k_2,-k_3)
 \epsilon_{\mu}(k_1) \epsilon_{\nu}(k_2) \epsilon_{\rho}(k_3).
\label{ampl}
\end{eqnarray}
The tensor ${\Theta^{\mu\nu\rho}_3}(a;k_1,k_2,k_3)$ is given by
\begin{eqnarray}
{\Theta^{\mu\nu\rho}_3}(a;k_1,k_2,k_3)&=&
-\,(k_1 \theta k_2)\,
\label{ampli}\\
&\times &
[(k_1-k_2)^\rho g^{\mu \nu} +(k_2-k_3)^\mu g^{\nu \rho} + (k_3-k_1)^\nu g^{\rho \mu}]
\nonumber \\
& -&
\theta^{\mu \nu}\,
[ k_1^\rho \, (k_2 k_3) - k_2^\rho \, (k_1 k_3) ]
\nonumber \\
& -&
\theta^{\nu \rho}\,
[ k_2^\mu \, (k_3 k_1) - k_3^\mu \, (k_2 k_1) ]
\nonumber \\
& -&
\theta^{\rho \mu}\,
[ k_3^\nu \, (k_1 k_2) - k_1^\nu \, (k_3 k_2) ]
\nonumber \\
& +& 
(\theta k_2)^\mu \,\left[g^{\nu \rho}\, k_3^2 - k_3^\nu k_3^\rho\right]
+(\theta k_3)^\mu\,\left[g^{\nu \rho}\, k_2^2 - k_2^\nu k_2^\rho\right]
\nonumber \\
& +& 
(\theta k_3)^\nu \,\left[g^{\mu \rho}\, k_1^2 - k_1^\mu k_1^\rho \right]
+(\theta k_1)^\nu \,\left[g^{\mu \rho}\, k_3^2 - k_3^\mu k_3^\rho \right]
\nonumber \\
& +& 
(\theta k_1)^\rho \,\left[g^{\mu \nu}\, k_2^2 - k_2^\mu k_2^\nu \right]
+(\theta k_2)^\rho \,\left[g^{\mu \nu}\, k_1^2 - k_1^\mu k_1^\nu \right]
\nonumber \\
& +& 
\theta^{\mu\alpha}(ak_1+k_2+k_3)_{\alpha} \,\left[g^{\nu \rho}\,(k_3 k_2)-k_3^\nu k_2^\rho\right]
\nonumber \\
& +& 
\theta^{\nu\alpha} (k_1+ak_2+k_3)_{\alpha} \,\left[g^{\mu \rho}\,(k_3 k_1)-k_3^\mu k_1^\rho\right]
\nonumber \\
& +& 
\theta^{\rho\alpha} (k_1+k_2+ak_3)_{\alpha} \,\left[g^{\mu \nu}\,(k_2 k_1)-k_2^\mu k_1^\nu\right]
\, ,
\nonumber 
\end{eqnarray}
where the 4-momenta $k_1,k_2,k_3$ are taken to be incoming, satisfying the momentum
conservation $(k_1+k_2+k_3=0)$. 
In  (\ref{ampli}) the freedom parameter $a$ 
appears symmetric in physical gauge bosons which enter the interaction point,
as one would expect.  
The amplitude (\ref{ampl}), for $a=3$, 
with the Z boson at rest gives the total rate
for the $Z \rightarrow \gamma\gamma$ decay:
\begin{eqnarray}
\Gamma_{Z\rightarrow \gamma\gamma}
&=&\frac{\alpha}{4} \;\frac{M^5_Z}{\Lambda^4_{\rm NC}}\; \sin^2 2\theta_W
{\rm K}^2_{Z\gamma \gamma}
\big({\vec E}_{\theta}^2 + {\vec B}_{\theta}^2\big),
\label{eqn3}
\end{eqnarray}
where ${\vec E}_{\theta}=\{{\theta}^{01},{\theta}^{02},{\theta}^{03}\}$ and
${\vec B}_{\theta}=\{{\theta}^{23},{\theta}^{31},{\theta}^{12}\}$
are dimensionless coefficients of order one, representing the time-space and space-space
noncommutativity, respectively. For the $Z$ boson at
rest, polarized in the direction of the third axis, we obtain the
following {\it polarized} partial width:
\be
\Gamma^3_{Z\rightarrow \gamma\gamma} =\frac{\alpha}{60}
\frac{M^5_Z}{\Lambda^4_{\rm NC}} \sin^2 2\theta_W {\rm
K}^2_{Z\gamma \gamma}
\left({\vec E}_{\theta}^2 + {\vec B}_{\theta}^2 + 42\left(
(\theta^{03})^2+(\theta^{12})^2\right)\right)\,. 
\label{eqn1p}
\ee
In order to estimate the scale of noncommutativity $\Lambda_{\rm NC}$ from 
$\Gamma_{Z\rightarrow \gamma\gamma}$,we 
consider new experimental possibilities at LHC.
According to the CMS Physics Technical Design Report \cite{CMS1},
around $10^7\;\, Z\to e^+ e^-$ events are expected to be recorded with $10 \;\,fb^{-1}$ of the data.
From this one can estimate the expected number of $Z \to \gamma\gamma$ events per $10 \;\,fb^{-1}$.
Assuming that $BR(Z \to \gamma\gamma) \sim 10^{-8}$ and using $BR(Z \to e^+e^-) = 3 \times 10^{-2}$,
we may expect to have $\sim 3$ events of $Z \to \gamma\gamma$ with $10\; fb^{-1}$.
Now the question is: What would be the background from $Z \to e^+e^-$ when the electron radiates
a very high-energy bremsstrahlung photon in the beam pipe or in the first layer(s) of the Pixel
Detector and is thus lost for the tracker reconstruction?
In that case, the electron would not be reconstructed and would be misidentified as a
photon. The probability of such an event should be evaluated from the
full detector simulation. According to the CMS note \cite{CMS2} which studies the $Z \to e^+e^-$
background for $Higgs \to \gamma\gamma$, the probability
to misidentify the electron as a photon is huge (see Fig. 3 in \cite{CMS2}) but
the situation can be improved by applying more stringent selections to the photon candidate when 
searching for $Z \to \gamma\gamma$ events \cite{Nikitenko}. However, the  irreducible
di-photon background (Fig. 3 in \cite{CMS2}) might also kill the signal.
In that case, one can only set the upper limits to the scale of noncommutativity from
the $Z \to \gamma\gamma$ rate.

In accord with the analysis of the LHC experimental expectations \cite{CMS1,CMS2,Nikitenko} it
is bona fide reasonable to assume that the lower bound for the branching ratio is
$BR(Z \rightarrow \gamma\gamma) \stackrel{<}{\sim} 10^{-8}$.
Next, choosing the lower central value of $|K_{Z\gamma\gamma}|=0.05$, from
the figures and the Table in \cite{Goran}, we find that
the upper bound to the scale of noncommutativity is 
$\Lambda_{\rm NC}\stackrel{>}{\sim}1.0\;\,\rm TeV$
for ${\vec E}_{\theta}^2 + {\vec B}_{\theta}^2\simeq 1$.
The obtained bound is strongly supported in \cite{Martin:2006gw}.

Clearly, the measurement of the $Z \rightarrow \gamma\gamma$ decay branching ratio
would fix the quantity $|K_{Z\gamma\gamma}/\Lambda_{\rm NC}^2|$, while
the inclusion of other triple gauge boson
interactions through $2\to 2$ scattering experiments \cite{Ohl:2004tn} would sufficiently
reduce the available parameter space of our model by more precisely determining
the relations among the couplings
${\rm K}_{\gamma\gamma\gamma}$, ${\rm K}_{Z \gamma\gamma}$,
${\rm K}_{ZZ\gamma}$, ${\rm K}_{ZZZ}$, ${\rm K}_{WW\gamma}$, and ${\rm K}_{WWZ}$.
Next, we summarize our results and compare with those obtained previously.

The first $Z\to\gamma\gamma$ calculation \cite{Mocioiu:2000ip} was performed
within a different model which  has different symmetries 
in comparison with ours and, because of the absence of 
the SW map, the model does not possess the commutative gauge invariance. 
Also, the  $Z\to\gamma\gamma$ rate obtained in \cite{Mocioiu:2000ip}
by imposing the  unitarity of the theory in the usual manner,
$\theta^{0i} = 0$, \cite{Seiberg:2000gc,Gomis:2000zz}, vanishes 
\footnote{The condition of unitarity can be covariantly generalized to
$\theta_{\mu\nu}\theta^{\mu\nu}= 2({\vec B}_{\theta}^2-{\vec E}_{\theta}^2)>0$ 
\cite{Carroll:2001ws}.}.

The partial width for the same process was obtained in \cite{Goran}
in the framework of similar theories, which, however, were not renormalizable. The present
results for the partial widths $\Gamma_{Z\rightarrow \gamma\gamma}$ and 
$\Gamma^3_{Z\rightarrow \gamma\gamma}$ are about three times
larger than those in \cite{Goran} and consistently symmetric with respect 
to time-space  and space-space noncommutativity.
In the polarized rate (\ref{eqn1p}) the third components 
($(\theta^{03})^2+(\theta^{12})^2$)
are enhanced relative to the other two components by a large factor, as expected.    
Also, the rate (\ref{eqn1p}) is enhanced 
by a factor of $\sim$ 3 with respect to the total rate (\ref{eqn3}). 
The upper limit to the scale of noncommutativity $\Lambda_{\rm NC}\stackrel{>}{\sim} 1$ 
TeV is significantly higher than in \cite{Goran}.
This bound is now firmer owing
to the regular behavior of the triple gauge boson
interactions (\ref{L1}-\ref{L2})
with respect to the one-loop renormalizability  of the NCSM
gauge sector.

After 10 years of the LHC running the integrated luminosity is expected
to reach $\sim 1000 \;\,fb^{-1}$,  \cite{CMS2}.
This means that for the assumed $BR(Z \to \gamma\gamma) \sim 10^{-8}$ we should have $\sim 300$
events of $Z \to \gamma\gamma$, that is we should be well above the background. On the other hand,
this result can also be understood as $\sim 3$ events with the $BR(Z \to \gamma\gamma) \sim 10^{-10}$,
which lifts the scale of noncommutativity up by a factor of $\sim 3$.
Therefore, with a more stringent selection of photon candidates and 
if the irreducible di-photon contamination becomes controllable,
the $Z \rightarrow \gamma\gamma$ decay will become a clean signature of
space-time noncommutativity in LHC experiments. 

Finally, the results of [17,18], while strongly supporting this computations, 
might also hint at the existence of new interaction symmetry of 
the noncommutative  gauge sector. 
Such new symmetry could be a responsible for the
renormalizability of the noncommutative matter sector including fermions and, next, 
for the main goal, i.e. in general, the physical realization of the Lorentz
invariance breaking at very high energies, respectively.


%

%

\end{document}